\documentclass[sigconf]{acmart}

\AtBeginDocument{%
  }

\usepackage{multicol}
\usepackage{multirow}
\usepackage{float}
\usepackage{bm}
\usepackage{algorithm}
\usepackage{algpseudocode}

\copyrightyear{2023}
\acmYear{2023}
\setcopyright{acmlicensed}
\acmConference[MM '23]{Proceedings of the 31st ACM International Conference on Multimedia}{October 29-November 3, 2023}{Ottawa, ON, Canada}
\acmBooktitle{Proceedings of the 31st ACM International Conference on Multimedia (MM '23), October 29-November 3, 2023, Ottawa, ON, Canada} 
\acmPrice{15.00}
\acmDOI{10.1145/3581783.3612272} 
\acmISBN{979-8-4007-0108-5/23/10}

\begin{document}

\title{Elucidate Gender Fairness in Singing Voice Transcription}
\author{Xiangming Gu}
\affiliation{%
  \institution{Integrative Sciences and Engineering Programme, NUS Graduate School, National University of Singapore}
  \country{Singapore}
  }
\email{xiangming@u.nus.edu}

\author{Wei Zeng}
\affiliation{%
  \institution{Integrative Sciences and Engineering Programme, NUS Graduate School, National University of Singapore}
  \country{Singapore}
  }
\email{w.zeng@u.nus.edu}

\author{Ye Wang}
\affiliation{%
  \institution{School of Computing, National University of Singapore}
  \country{Singapore}
}
\email{wangye@comp.nus.edu.sg}

\begin{abstract}
It is widely known that males and females typically possess different sound characteristics when singing, such as timbre and pitch, but it has never been explored whether these gender-based characteristics lead to a performance disparity in singing voice transcription (SVT), whose target includes pitch. Such a disparity could cause fairness issues and severely affect the user experience of downstream SVT applications. Motivated by this, we first demonstrate the female superiority of SVT systems, which is observed across different models and datasets. We find that different pitch distributions, rather than gender data imbalance, contribute to this disparity. To address this issue, we propose using an attribute predictor to predict gender labels and adversarially training the SVT system to enforce the gender-invariance of acoustic representations. Leveraging the prior knowledge that pitch distributions may contribute to the gender bias, we propose conditionally aligning acoustic representations between demographic groups by feeding note events to the attribute predictor. Empirical experiments on multiple benchmark SVT datasets show that our method significantly reduces gender bias (up to more than $50\%$) with negligible degradation of overall SVT performance, on both in-domain and out-of-domain singing data, thus offering a better fairness-utility trade-off.
\end{abstract}





\maketitle
\section{Introduction}

\begin{figure}[h]
\centering
\vskip -0.1in
\includegraphics[width=0.9\linewidth]{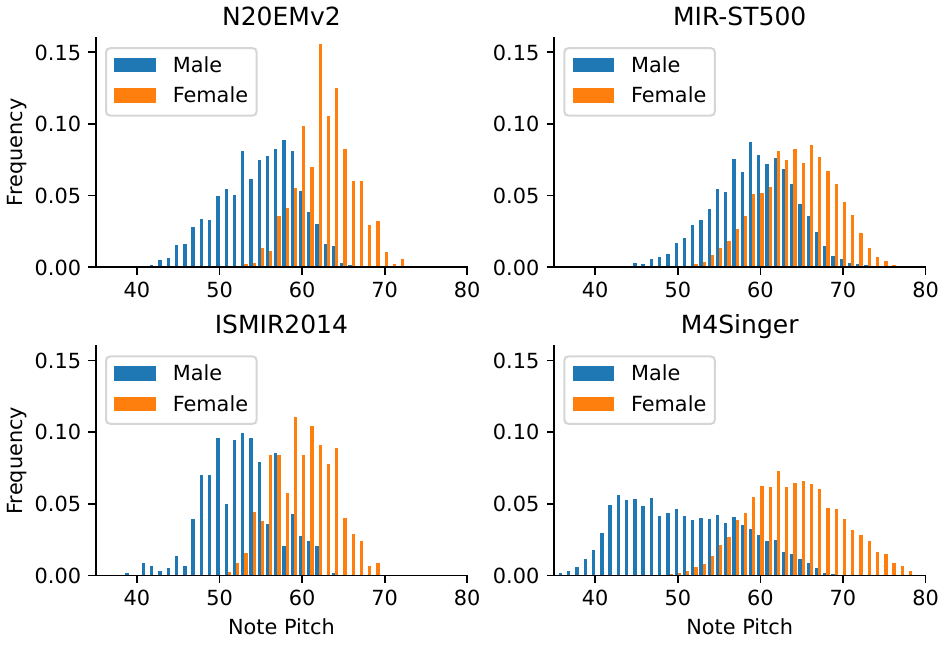}
\caption{Pitch distributions of various singing voice transcription datasets.}
\vskip -0.2in
\label{fig_pitch}
\end{figure}

Singing is a rich audio signal that consists of information from two aspects: textual and musical. The textual modality pertains to the lyrics, while the musical modality encompasses note events. Automatic lyric transcription (ALT) can be used to retrieve lyrical/textual data, as evidenced by prior literature \cite{dabike2019automatic,demirel2020automatic,demirel2021mstre,zhang2021pdaugment, gu2022mm,ou2022transfer,gao2022genre,gao2022automatic}. On the other hand, singing voice transcription (SVT) is utilized to retrieve note events, which include onsets, offsets, and pitches \cite{mauch2015computer,fu2019hierarchical, hsu2021vocano,wang2021preparation,kum2022pseudo,wang2022musicyolo,gu2023deep}. Transcribed information can enable the development of singing voice synthesis \cite{ren2020deepsinger, liu2022diffsinger} and aid in education \cite{yang2022multi, molina2013fundamental, gupta2017perceptual} and therapy \cite{tam2007movement}. It is important to note that males and females have distinct sound characteristics in their singing voices \cite{latinus2012discriminating}, e.g. timbre. Additionally, compared to lyrics, the note events tend to be more subject to explicit gender-related biases, particularly for pitch, as shown in prior literature \cite{puts2007men, simpson2009phonetic} that males tend to have lower average pitch than females. This is also consistent with our analysis of four SVT datasets, namely N20EMv2 \cite{gu2023deep}, MIR-ST500 \cite{wang2021preparation}, ISMIR2014 \cite{molina2014evaluation}, and M4Singer \cite{zhang2022m4singer}. As shown in Fig. \ref{fig_pitch}, we observe that the pitch range of females is generally higher than that of males across these four SVT datasets. Besides, the proportion of males and females for each pitch value is also different. Consequently, a critical \textbf{fairness} question arises: since SVT systems target pitches, will the performance of these systems favor one gender over the other? Before we delve into this question, it is imperative to elucidate why this question holds substantial significance.

The rapid progress in machine learning techniques has facilitated their successful integration into various downstream applications, streamlining decision-making processes and reducing the need for repeated human efforts. However, the presence of bias in machine learning systems can lead to the discriminatory treatment of certain groups with sensitive attributes such as gender, age, and race, resulting in unfair decisions. Moreover, general usage of biased machine learning systems can reinforce the stereotypes and exclude certain groups from the opportunities. This phenomenon is not only limited to traditional decision-making scenarios such as loan applications, hiring, legal proceedings, and policy-making \cite{zemel2013learning}, but also frequently appears in recent deep learning applications, including visual recognition \cite{wang2020towards, sarhan2020fairness, tartaglione2021end}, natural language processing \cite{guo2022auto, ding2022word}, speech processing and recognition \cite{rajan2022aequevox, dheram2022toward, liu2022towards, gorrostieta2019gender, wagner2023dawn, fenu2020exploring, chen2022exploring}, recommendation systems \cite{sonboli2021fairness, li2021user}, and generative models \cite{choi2020fair, friedrich2023fair}. Consequently, the issue of fairness has gained prominence in the machine learning community due to its pervasive nature and potential societal consequences. Returning to our SVT task, the unfairness of systems could directly lead to user inconvenience and negatively impact their experience. Applying biased SVT systems to downstream applications could cause more fairness issues. We can consider an automatic sight-singing exercise \cite{yang2022multi}, where an SVT system could serve as an intermediate stage to transcribe the singing voice into musical notes. However, if a gender-biased SVT system is utilized, it could result in an unsatisfying user experience for certain groups, as well as reinforce gender stereotypes, which would be unfair to individuals of different genders.

In this study, we elucidate the fairness issue in the field of SVT. Our investigation demonstrates that SVT systems tend to perform better on females compared to males, raising questions about their fairness. Then, we assume that this gender bias is attributable to the differences of singing voices, especially in terms of pitch distributions, in different demographic groups. We utilize self-supervised-learning (SSL)-based SVT systems \cite{gu2023deep}, which represent the current state-of-the-art in this field, to develop a bias mitigation approach. In contrast to previous research, which normally focuses on the in-domain fairness, we also pay attention to the out-of-domain fairness as user data is sometimes sampled from out-of-domain distribution. To implement this, our approach adopts the adversarial learning framework to ``unlearn'' gender-related information in the acoustic representations. Considering the effects of pitch distributions, we propose a note-conditional attribute predictor to conditionally align the representations between female and male groups by conditioning on note events. Empirical results from various SVT datasets confirm the effectiveness of our approach in mitigating gender bias. Our contributions are summarized as below:
\begin{itemize}
    \item We provide evidence and analysis of the prevalence and source of gender bias in SVT systems. To the best of our knowledge, this is the first attempt at fairness in singing-centric deep learning.
    \item We first introduce a note-conditioned adversarial learning approach to achieve fair representation learning in audio modality, resulting in a significant reduction in the performance gap between the two gender groups on various benchmark SVT datasets while maintaining a good fairness-utility trade-off. Our method is effective in reducing biases on both in-domain and out-of-domain data.
    \item We demonstrate the superiority of our note-conditioned adversarial learning approach through comparisons with baseline adversarial learning and domain-independent training.
\end{itemize}

\section{Related Work}

\subsection{Singing Voice Transcription}
Singing voice transcription (SVT) involves various sub-tasks, including pitch estimation and onset/offset detection. Earlier approaches \cite{mauch2014pyin, mauch2015computer, yang2017probabilistic, nishikimi2020bayesian} typically relied on statistical models, such as Bayesian models and Hidden Markov Models (HMM), to predict fundamental frequency (F0) and note segmentation. In contrast, more recent SVT methods \cite{fu2019hierarchical, hsu2021vocano, kum2022pseudo} have predominantly employed deep learning techniques, such as CNN and LSTM, and have demonstrated superior performance. Despite the promising SVT performance achieved by these methods, the intrinsic difficulty of curating large-scale, high-quality SVT datasets presents an obstacle to further improvements. To address this challenge, several approaches have been proposed. Among them, VOCANO \cite{hsu2021vocano} employed the Virtual Adversarial Training (VAT) \cite{miyato2018virtual} to train the note segment network on both labeled and unlabeled data. In \cite{kum2022pseudo}, pseudo labels are obtained by quantizing frame-level pitch contours for training on unlabeled audio data. MusicYOLO \cite{wang2022musicyolo} utilized the object detection model YOLOX \cite{ge2021yolox}, which has been trained on the image domain, to locate notes in the audio spectrogram. Recently, \cite{gu2023deep} adapted self-supervised learning (SSL) models from the speech domain to the SVT task, thus alleviating the label insufficiency as well as achieving state-of-the-art SVT performance. 

\subsection{Fairness and Bias Mitigation}
The notion of fairness in machine learning systems is defined as the absence of any discrimination based on sensitive attributes when making decisions. It can be categorized into group fairness \cite{hardt2016equality, zhao2019conditional} and individual fairness \cite{dwork2012fairness, kusner2017counterfactual}. The former requires that there are no disparities among different demographic groups \cite{zhao2019conditional}, while the latter requires that similar individuals receive similar predictions \cite{kusner2017counterfactual}. In our work, we focus on group fairness, which can be assessed by criteria, including independence, separation, and sufficiency \cite{barocas-hardt-narayanan, shen2022fair}, along with metrics such as demographic parity, equalized odds, equal opportunity, and accuracy parity \cite{verma2018fairness, zhao2019conditional, mehrabi2021survey}.

Numerous approaches have been proposed to mitigate bias in machine learning systems. Among them, adversarial learning has emerged as a powerful technique for removing sensitive attributes in the representations used for prediction. For instance, \cite{louppe2017learning} repurposed the framework of generative adversarial networks (GANs) \cite{goodfellow2020generative} to satisfy demographic parity. \cite{madras2018learning} used an adversary to predict sensitive attributes from latent representations and a decoder to reconstruct the input data from the latent representations and predicted attributes in adversarial learning. Additionally, they proposed different adversarial objectives according to the target group fairness criteria. \cite{kim2019learning} proposed using a bias prediction network to minimize mutual information between latent representations and bias through adversarial learning. In contrast to these approaches, \cite{zhao2019conditional} proposed conditional alignment of latent representations to strike a better balance between fairness and utility. In addition to adversarial learning, alternative methods have been proposed to mitigate bias. \cite{wang2020towards} advocated domain independent training, where different classifiers are trained for different demographic groups. \cite{sarhan2020fairness, tartaglione2021end} disentangled the latent representations for task predictions and sensitive attributes, respectively, to reduce the influence of the sensitive attributes on the model's decision-making process.

\subsection{Adversarial Learning for Invariance}
In addition to the adversarial learning framework used in the fairness community, our work also shares similarities with unsupervised domain adaptation, which aims to learn domain-invariant representations. \cite{ganin2015unsupervised} proposed a domain classifier to learn the representations that are invariant to the domain shift, with an adversary implemented by a gradient reverse layer. \cite{tzeng2017adversarial} used a discriminator to align the distributions of source domain and target domain in the representation space, following the training of GANs \cite{goodfellow2020generative}. Similar to \cite{zhao2019conditional}, \cite{zhao2017learning, long2018conditional} proposed a conditional discriminator that takes into account the multimodal nature of feature distributions. One key difference compared to the fairness scenario is that labels in the target domain are not available in domain adaptation. Therefore, the conditions in \cite{zhao2017learning, long2018conditional} are the label predictions, rather than the ground-truth labels used in \cite{zhao2019conditional}. 

\section{Preliminary for SVT}\label{preliminary}

We recap the formulation and solutions of singing voice transcription (SVT), which is defined according to the framework introduced in \cite{wang2021preparation, gu2023deep}. The input to an SVT system is the waveform $\bm{x}$ while the output $\bm{y}=[(o_1, f_1, p_1), ..., (o_n, f_n, p_n),...,$ $(o_N, f_N, p_N)]$ is a sequence of note events, where $o_n, f_n, p_n$ represent the onset/offset/pitch of each note, respectively. Consequently, the task of SVT can be regarded as a sequence-to-sequence problem. Since it is challenging to supervise the entire model using the ground truth note events directly, frame-level labels $\bm{O}, \bm{S}, \bm{V}, \bm{P}$ are constructed to mark the onset/silence/octave class/pitch class of each frame. Specifically, $\bm{O}, \bm{S}$ comprise binary classes, whereas $\bm{V}$ and $\bm{P}$ have multiple classes. The number of categories of pitch classes is fixed to 12, while the number of categories of octaves are chosen based on the pitch range. For example, the octave class of C4 is 4 and the pitch class is C. We add an additional octave/pitch class to represent the pitch of silence. The loss function is formulated to minimize the empirical risk as follows:
\begin{align}\label{eq1}
    \mathcal{L}_{\text{SVT}}(\hat{\bm{y}}, \bm{y})=\frac{1}{T}\sum_{t=1}^T[&l_{\text{BCE}}(\hat{O}_t, O_t)+l_{\text{BCE}}(\hat{S}_t, S_t)+\\\nonumber
    &l_{\text{CE}}(\hat{V}_t, V_t)+l_{\text{CE}}(\hat{P}_t, P_t)],
\end{align}where $T$ refers to the number of frames, $\hat{O}_t, \hat{S}_t, \hat{V}_t, \hat{P}_t$ denote the frame-level predictions while $\hat{\bm{y}}$ represents note-level predictions. After post-processing, $\hat{O}_t, \hat{S}_t, \hat{V}_t, \hat{P}_t$ are transformed to $\hat{\bm{y}}$. Interested readers can refer to \cite{wang2021preparation, gu2023deep} for more detailed information. An SVT system consists of an acoustic encoder and a note predictor. The note predictor is a simple linear layer, while the design of acoustic encoder can vary. For instance, \cite{wang2021preparation} employed EfficientNet \cite{tan2019efficientnet}, while \cite{gu2023deep} adapted self-supervised-learning models, e.g. wav2vec 2.0 as the acoustic encoder. To evaluate SVT systems, f1-scores of COnPOff (correct onset, pitch, and offset), COnP (correct onset and pitch), and COn (correct onset) are commonly used. These metrics were proposed in \cite{molina2014evaluation} and have since been widely adopted. The first two metrics are related to the accuracy of pitch estimation. We use the default tolerances as implemented in the python package \textit{mir\_eval} \cite{raffel2014mir_eval} for the evaluation of SVT systems in this work.

\begin{table*}[h]
\caption{COnPOff/COnP/COn F1-score (\%) of state-of-the-art SVT systems in \cite{gu2023deep} on multiple datasets. We using \textbf{bold face} to highlight the gender group with better performance and \textcolor{red}{\textbf{red bold face}} to mark the results with large bias.}
\begin{tabular}{l|c|cccc|cccc|cccc}
\toprule
\multirow{2}{*}{Dataset} & \multirow{2}{*}{Model} & \multicolumn{4}{c|}{COnPOff (\%)}           & \multicolumn{4}{c|}{COnP (\%)}              & \multicolumn{4}{c}{COn (\%)}               \\ \cline{3-14} 
                  &       & total $\uparrow$ & female $\uparrow$         & male $\uparrow$  & gap   & total $\uparrow$ & female $\uparrow$         & male $\uparrow$  & gap  & total $\uparrow$ & female $\uparrow$         & male $\uparrow$  & gap  \\ \midrule 
\multirow{3}{*}{MIR-ST500}  &  model1     & 52.39 & \textbf{53.11} & 51.47 & -1.64  & 70.73 & \textbf{72.54} & 68.42 & -4.12  & 78.32 & \textbf{79.36} & 77.00 & -2.36 \\ 
   & model2        & 34.55 & \textbf{35.29} & 33.60 & -2.55 & 51.64 & \textbf{52.76} & 50.21 & -2.55 & 71.33 & \textbf{72.07} & 70.39 & -1.68 \\
   & model3        & 52.84 & \textbf{54.02} & 51.33 & -2.69  & 70.00 & \textbf{71.17} & 67.85 & -3.32  & 78.05 & \textbf{78.84} & 77.05 & -1.79 \\
\midrule
\multirow{3}{*}{N20EMv2}   &  model1          & 55.20 & \textbf{60.96} & 51.55 & \textcolor{red}{\textbf{-9.41}}  & 72.03 & \textbf{79.76} & 67.11 & \textcolor{red}{\textbf{-12.65}}  & 88.51 & \textbf{90.63} & 87.16 & -3.47 \\ 
     & model2        & 68.62 & \textbf{74.82} & 64.68 & \textcolor{red}{\textbf{-10.14}} & 75.69 & \textbf{81.27} & 72.14 & \textcolor{red}{\textbf{-9.12}} & 92.83 & \textbf{94.33} & 91.88 & -2.44 \\
     & model3        & 73.06 & \textbf{78.34} & 69.69 & \textcolor{red}{\textbf{-8.65}}  & 79.56 & \textbf{84.38} & 76.49 & \textcolor{red}{\textbf{-7.89}}  & 93.66 & \textbf{94.96} & 92.83 & -2.13 \\ 
\midrule
\multirow{3}{*}{ISMIR2014}  &  model1              & 52.58 & \textbf{61.22} & 45.28 & \textcolor{red}{\textbf{-15.94}} & 67.75 & \textbf{74.90} & 61.70 & \textcolor{red}{\textbf{-13.20}} & 92.13 & 91.93 & \textbf{92.30} & +0.37 \\ 
   & model2            & 57.35 & \textbf{62.42} & 53.06 & \textcolor{red}{\textbf{-9.36}} & 72.15 & \textbf{79.35} & 66.06 & \textcolor{red}{\textbf{-13.29}} & 91.53 & \textbf{92.25} & 90.92 & -1.33 \\
   & model3            & 59.95 & \textbf{65.61} & 55.16 & \textcolor{red}{\textbf{-10.45}} & 73.85 & \textbf{80.55} & 68.19 & \textcolor{red}{\textbf{-12.36}} & 92.80 & \textbf{93.18} & 92.49 & -0.69 \\ 
\midrule
\multirow{4}{*}{M4Singer} & wav2vec2 & 53.66 & \textbf{57.27}& 49.93& \textcolor{red}{\textbf{-7.34}}& 61.95& \textbf{66.36}& 57.38& \textcolor{red}{\textbf{-8.98}}& 82.60& 81.66& \textbf{83.58}& +1.91 \\
& Hubert & 55.11 & \textbf{58.39} & 51.73 & \textcolor{red}{\textbf{-6.66}} &  64.17 & \textbf{68.10} & 60.10 & \textcolor{red}{\textbf{-8.00}} & 82.13 & 81.53 & \textbf{82.76} & +1.23 \\
& wavLM & 57.06 &  \textbf{60.33} & 53.68 & \textcolor{red}{\textbf{-6.66}} & 65.40 & \textbf{68.87} & 61.82 & \textcolor{red}{\textbf{-7.05}} & 82.73 & 82.29 & \textbf{83.19} & +0.90  \\
& data2vec & 53.98 & \textbf{57.45} & 50.40 & \textcolor{red}{\textbf{-7.05}} & 63.04 & \textbf{67.09} & 58.86 & \textcolor{red}{\textbf{-8.23}} & 82.30 & 82.13 & \textbf{82.48} & +0.35  \\   
\bottomrule
\end{tabular}
\label{tbl_bias}
\vskip -0.1 in
\end{table*}
\section{Fairness Analysis for SVT}
\subsection{Female Superiority in SVT Performance: Evidence from Multiple Datasets}\label{gap_data}

We evaluate the performance of state-of-the-art singing voice transcription (SVT) systems from \cite{gu2023deep} on three benchmark datasets, including MIR-ST500 \cite{wang2021preparation}, N20EMv2 \cite{gu2023deep}, and ISMIR2014 \cite{molina2014evaluation}. The results for two gender groups are presented in Table \ref{tbl_bias}, where ``model1'' is trained on the MIR-ST500 training split, ``model2'' on the N20EMv2 training split, and ``model3'' on both training sets. These models were built based on wav2vec 2.0 \cite{baevski2020wav2vec}. The ISMIR2014 dataset already has the gender label for each song. The gender labels for the N20EMv2 and MIR-ST500 can be obtained by directly listening to the audio recordings. For double confirmation, we also check the video modality provided in N20EMv2 and the original Youtube links in MIR-ST500. Across all datasets, we observe that the SVT performances of three models on females consistently outperform that on males in terms of COnPOff and COnP f1-scores, which are the metrics related to pitch estimation. However, the COn performance does not demonstrate the consistent female superiority. We compute the performance gap as the difference between male and female metrics, i.e., $\text{metric}_{\text{gap}}=\text{metric}_{\text{male}}-\text{metric}_{\text{female}}$. We find that the performance gaps between gender groups are significant for pitch-related metrics on the N20EMv2 and ISMIR2014 datasets. Even though the performance gap is comparably small for MIR-ST500, female superiority in SVT performance remains valid. 

Apart from the datasets evaluated in \cite{gu2023deep}, we conduct experiments on a recent Mandarin singing dataset called M4Singer \cite{zhang2022m4singer}. This dataset comprises data from 20 singers across four main voice types (soprano, alto, tenor, and bass). Since M4Singer lacks official training-test splits, we manually partition the data, selecting data from two male and two female singers representing the above four voice types for the test split and the data from the remaining singers for the training split. To ensure a fair test split, we make sure that the total duration of selected female data in the test split is almost equal to that of male data. We follow the training configuration in \cite{gu2023deep} to train our SVT system from scratch using wav2vec 2.0, making minimal modifications to achieve high SVT performance. We add an additional octave category to the classifier to account for the larger pitch range of M4Singer. The M4Singer dataset already provides the gender labels, similar to ISMIR2014. The results in Table \ref{tbl_bias} show that the SVT performance of females is still better than that of males, with a significant performance gap.

\begin{figure}[h]
\centering
\includegraphics[width=\linewidth]{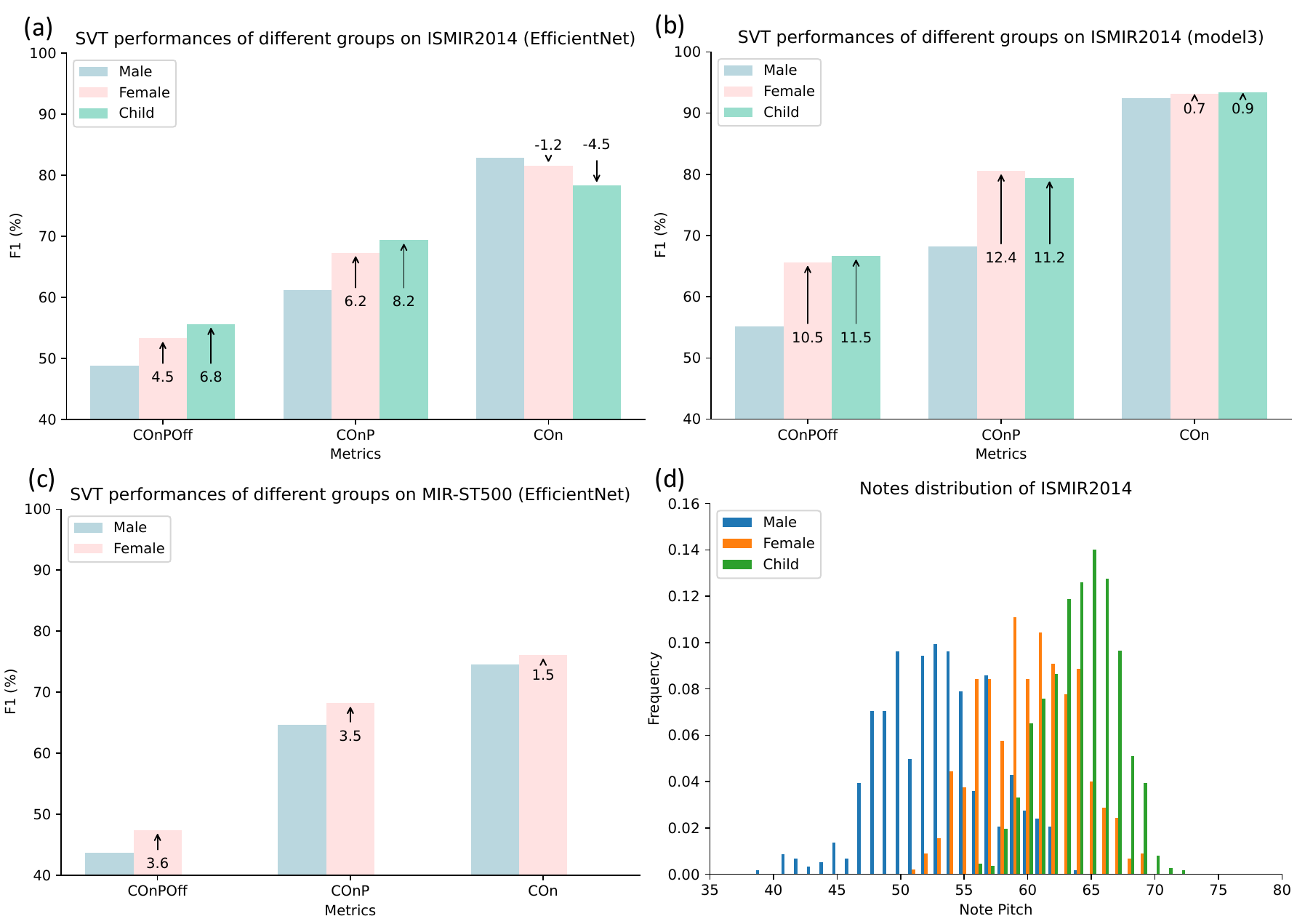}
\caption{(a) SVT performance of EfficientNet on ISMIR2014. (b) SVT performance of model3 on ISMIR2014. (c) SVT performance of EfficientNet on MIR-ST500. (d) Pitch distributions of male/female/child groups on ISMIR2014.} 
\label{fig_bias}
\vskip -0.2 in
\end{figure}

\subsection{Female Superiority in SVT Performance: Evidence from Multiple Models}\label{gap_model}
The models evaluated in Sec. \ref{gap_data} are all based on wav2vec 2.0 \cite{baevski2020wav2vec}. However, it is important to investigate whether the observed female superiority is limited to this particular model choice. To explore this question, we replace the wav2vec 2.0 trained on M4Singer with other self-supervised-learning (SSL) models, e.g. Hubert \cite{hsu2021hubert}, wavLM \cite{chen2022wavlm}, and data2vec \cite{baevski2022data2vec}. Although these models have different SSL objectives and slightly different model architectures, we find that the SVT performance of the female group is still better than the male group with significant margins, as shown in Table \ref{fig_bias}. To strengthen our findings, we further evaluate the performance of EfficientNet-based SVT system in \cite{wang2021preparation}. As presented in Fig. \ref{fig_bias} (a) and (c), we note that on the MIR-ST500 and ISMIR2014 datasets, the SVT performance of female group still outperforms that of the male group in terms of COnPOff and COnP f1-scores, which is consistent with our earlier conclusion. 

\subsection{Possible Reasons for Female Superiority}\label{gap_reason}

From Sec. \ref{gap_data} and Sec. \ref{gap_model}, we conclude that female superiority in SVT performance is valid across different datasets and model choices. To explore possible reasons behind this phenomenon, we first examine the statistics of the SVT datasets to investigate whether the singing datasets typically possess the property of gender data imbalance. As presented in Table \ref{tbl_stats}, we include the statistics of two gender groups in the training splits of MIR-ST500, N20EMv2, and M4Singer (ISMIR2014 is only used for evaluation). While MIR-ST500 has a larger proportion of female data, N20EMv2 and M4Singer have larger proportions of male data. Nevertheless, all the SVT systems trained on these datasets favored females, indicating that the source of bias is not merely the data imbalance. We hypothesize that the gender bias in SVT performance is attributed to the differences of sound characteristics across different demographic groups. Specifically, we assume that pitch distributions make substantial contributions to the female superiority. As shown in Fig. \ref{fig_pitch}, we find that (1) the female group generally has higher pitch range than the male group; (2) the proportion of male and female labels for each specific pitch is different. 

\begin{table}[t]
\caption{Demographic statistics of SVT training splits.}
\vskip -0.1in
\begin{tabular}{l|ccc|ccc}
\toprule
\multirow{2}{*}{Dataset} & \multicolumn{3}{c}{Songs Num} & \multicolumn{3}{|c}{Duration (h)} \\
                         & total    & female    & male   & total     & female    & male     \\
\midrule
MIR-ST500                & 400      & \textbf{221}       & 179    & 27.62     & \textbf{15.13}     & 12.49    \\
N20EMv2                  & 123      & \,\,52        & \textbf{\,\,71}     & \,\,6.44      & \,\,2.74      & \textbf{\,\,3.70}     \\
M4Singer                 & 521      & 246          & \textbf{275}        & 22.71     & 10.27     & \textbf{12.44}   \\
\bottomrule
\end{tabular}
\vskip -0.2in
\label{tbl_stats}
\end{table}

To further support our assumption, we perform an additional evaluation on the ISMIR2014 dataset, which includes the child group. Typically, children have different inherent properties in sound voices compared to female adults and male adults \cite{moore1991comparison, whiteside2000some}. For instance, the pitch distribution of the child group is different from both the female group and male group, as present in Fig. \ref{fig_bias} (d). Consequently, we find that both the EfficientNet-based SVT system and wav2vec 2.0-based SVT system (model3) demonstrate performance disparity among the three groups. As shown in Fig. \ref{fig_bias} (a) and (b), the SVT performance of the child group significantly outperforms that of the male group while is close to that of the female group. To interpret this, the pitch distribution difference between the male group and the child group is large while the difference between the female group and the child group is comparatively subtle. As there are no existing SVT annotations for child training data, we narrow down the scope of this work to gender fairness and leave the discussion on age fairness to future work. Similarly, as present in Table \ref{tbl_bias}, the performance gap on the MIR-ST500 dataset is not as large as the other three datasets. We assume the reason is that the singing recordings in MIR-ST500 were performed by professional singers, resulting in smaller pitch distribution difference between two gender groups. 

\begin{figure}[t]
\centering
\includegraphics[width=0.9\linewidth]{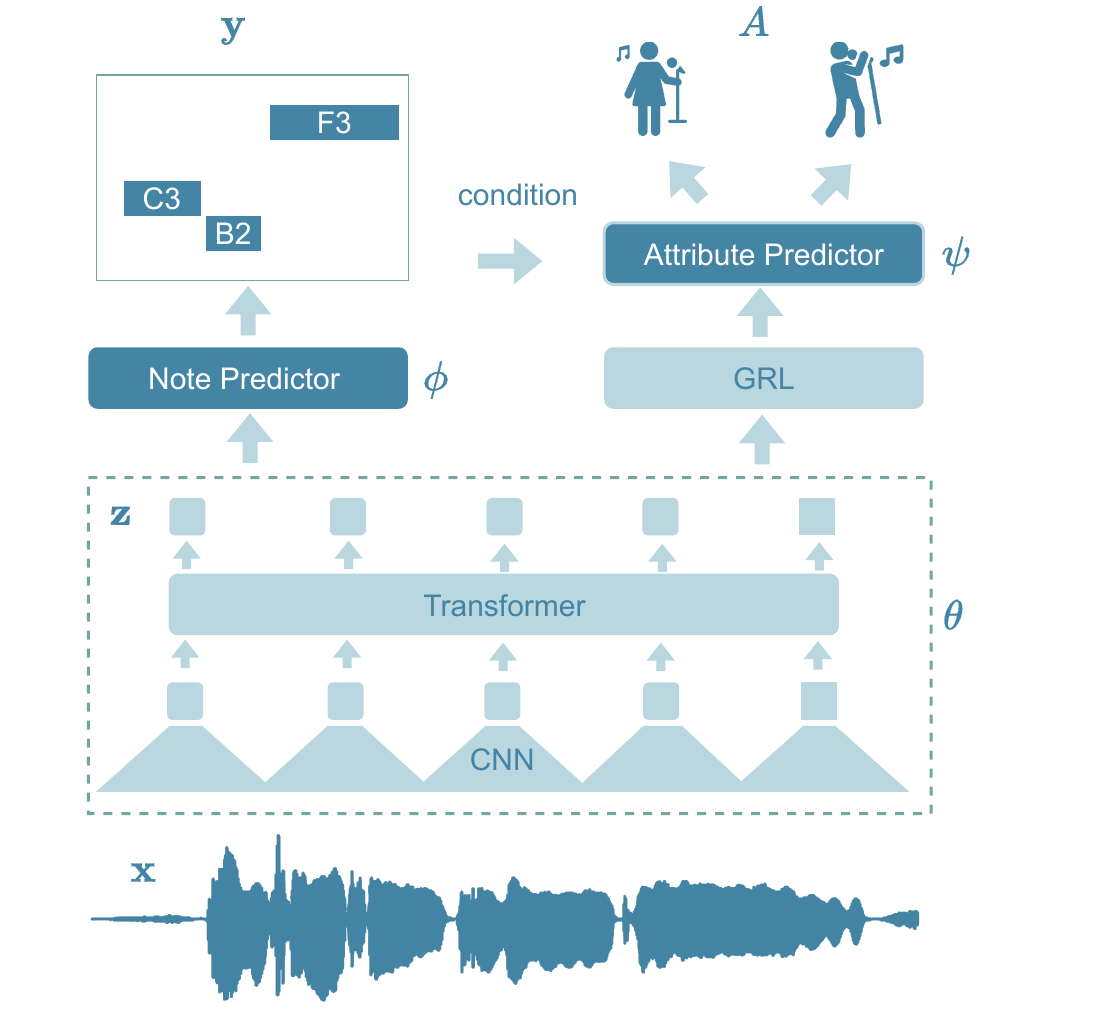}
\caption{Framework of note-conditioned adversarial learning for singing voice transcription.}
\label{fig_framework}
\vskip -0.2 in
\end{figure}

\section{Bias Mitigation for SVT}
\subsection{Problem Formulation and Basic Framework}
Suppose the training samples are drawn from the domain $\mathcal{D}_{\mathcal{S}}=\{(\bm{x}^{(n)}, \bm{y}^{(n)}, A^{(n)})\}_{n=1}^{N_{\mathcal{S}}}$, while the test data are sampled from the domain $\mathcal{D}_{\mathcal{T}}=\{(\bm{x}^{(n)}, \bm{y}^{(n)}, A^{(n)})\}_{n=1}^{N_{\mathcal{T}}}$. Here $\bm{x}^{(n)}$ represents the raw waveform of the singing audio, $\bm{y}^{(n)}$ represents the ground-truth note events, and $A^{(n)}$ is the sensitive attribute. In our study, $A=0$ represents female singers, while $A=1$ represents male singers.  As explained in Sec. \ref{preliminary}, the basic SVT system consists of two primary components: an acoustic encoder $\theta$ and a note predictor $\phi$. We select self-supervised-learning (SSL) models as our acoustic encoder due to their state-the-of-art performance in SVT tasks \cite{gu2023deep}. 

Fig. \ref{fig_framework} illustrates our proposed bias mitigation framework for the SVT task. The input singing waveform $\bm{x}$ is first processed by the acoustic encoder, which consists of a CNN and a transformer \cite{vaswani2017attention}. We refer readers to original papers of self-supervised-learning (SSL) models, e.g. wav2vec 2.0 \cite{baevski2020wav2vec}, Hubert \cite{hsu2021hubert}, wavLM \cite{chen2022wavlm}, and data2vec \cite{baevski2022data2vec} for details on these models, since they are not the focus of this paper. These models are first pre-trained on unlabeled speech data and then adapted to our singing data using a linear probing and then full fine-tuning approach \cite{gu2023deep}. After the acoustic encoder $\theta$, we obtain the acoustic features $\bm{z}\in \mathcal{R}^{T\times D}$, where $T$ is the number of frames and $D$ is the number of feature dimensions. The note predictor $\phi$ is parameterized by a linear layer, and accepts the acoustic features $\bm{z}$ to predict the frame-level labels: $\hat{O}_t, \hat{S}_t, \hat{V}_t, \hat{P}_t=\phi(\bm{z}_t)$. The loss function for SVT task in domain $\mathcal{D}_{\mathcal{S}}$ is:
\begin{equation}\label{eq2}
    \mathcal{L}_y=\mathbb{E}_{(\bm{x}, \bm{y})\sim\mathcal{D}_{\mathcal{S}}}\left[\mathcal{L}_{\text{SVT}}(\hat{\bm{y}}, \bm{y})\right],\,\, \hat{\bm{y}}=\phi(\bm{z})=\phi\circ\theta(\bm{x}),
\end{equation}where $\mathcal{L}_{\text{SVT}}$ is defined in Eq. \ref{eq1}. To evaluate the SVT performance, we compute the f1-scores of the COnPOff, COnP, and COn in domain $\mathcal{D}_{\mathcal{T}}$. We refer to these metrics as utility metrics:
\begin{equation}
    U=\mathbb{E}_{(\bm{x}, \bm{y})\sim\mathcal{D}_{\mathcal{T}}}\left[\text{metric}(\hat{\bm{y}}, \bm{y})\right],
\end{equation}where $\text{metric}$ can be the f1-score of COnPOff or COnP. As our objective is to mitigate gender bias in SVT systems, we propose the following fairness metrics, which focus on the performance disparity between two demographic groups: 
\begin{equation}
    F=\mathbb{E}_{(\bm{x}, \bm{y},A=1)\sim\mathcal{D}_{\mathcal{T}}}\left[\text{metric}(\hat{\bm{y}}, \bm{y})\right]-\mathbb{E}_{(\bm{x}, \bm{y},A=0)\sim\mathcal{D}_{\mathcal{T}}}\left[\text{metric}(\hat{\bm{y}}, \bm{y})\right],
\end{equation}
This definition is similar to the accuracy parity in \cite{zhao2019conditional}: $P(\hat{\bm{y}}\neq \bm{y}|A=0)=P(\hat{\bm{y}}\neq \bm{y}|A=1)$ only except that we replace accuracy with f1-score as the evaluation metric for parity. To achieve bias mitigation, we formulate the problem as a bundle of optimization objectives:
\begin{equation}\label{eq5}
    \left\{
    \begin{array}{l}
         \max_{\theta, \phi} \min\{F(\theta, \phi), 0\} \\
         \max_{\theta, \phi} U(\theta, \phi)
    \end{array}
    \right.
\end{equation}


\subsection{Adversarial Learning for Fairness}\label{AL}
The optimization objectives in Eq. \ref{eq5} cannot be directly optimized due to the unavailability of the test data $\mathcal{D}_{\mathcal{T}}$ during the training of $\theta, \phi$. To mitigate the gender bias, we propose an adversarial learning framework to learn fair acoustic representations by assuming that  the acoustic encoder cannot discriminate between two gender groups for SVT task. To achieve this, we employ an attribute predictor $\psi$ that predicts the labels of sensitive attributes, such as gender. The goal is to eliminate the gender information in the acoustic features $\bm{z}$ while preserving the information necessary for predicting the note events. The binary cross-entropy between gender predictions and ground truth gender labels serves as the learning objective of the attribute predictor:
\begin{equation}\label{eq6}
    \mathcal{L}_A=\mathbb{E}_{(\bm{x}, s)\sim\mathcal{D}_{\mathcal{S}}}\left[\frac{1}{T}\sum_{t=1}^Tl_{\text{BCE}}(\hat{A}_t, A)\right],\,\, \hat{A}=\psi(\bm{z})=\psi\circ\theta(\bm{x}).
\end{equation}To perform frame-level gender classification independently, each frame of acoustic features, denoted as $\bm{z}_t$, is annotated with the same attribute label $A$. We then average the frame-level classification loss to obtain the song-level or utterance-level loss. We assume that the temporal model structure in SSL models has sufficiently learned the gender representations and thus do not require a temporal attribute predictor. Our preliminary experiments show no empirical gains by including a temporal attribute predictor.

From a distribution alignment perspective, our objective is to achieve gender-invariant acoustic features, which requires that the distribution $P(\bm{z}|A=0)$ and the distribution $P(\bm{z}|A=1)$ be similar. The loss function in Eq. \ref{eq6} serves as a proxy measure for the distance between the two distributions, and thus, the attribute predictor $\psi$ must be trained to be powerful enough to distinguish between $P(\bm{z}|A=0)$ and $P(\bm{z}|A=1)$. Additionally, the acoustic encoder $\theta$ must be trained to deceive $\psi$. Therefore, the loss function for the acoustic encoder is formulated as $\mathcal{L}_y-\lambda\mathcal{L}_A$, where $\lambda$ is a hyper-parameter that balances the two loss terms. Theoretically, this can be implemented by a gradient reverse layer (GRL) \cite{ganin2015unsupervised}.  From the perspective of fairness criteria, the adversarial learning approach enforces that $\bm{z}\perp A$. Since $\hat{\bm{y}}=\phi(\bm{z})$, $\hat{\bm{y}}\perp A$ can be further enforced, which is the independence criteria \cite{barocas-hardt-narayanan}.

\subsection{Note-conditioned Adversarial Learning}
We have observed that there is a difference in pitch distributions between males and females, which may contribute to the performance disparity. Motivated by this, we propose the conditional distribution alignment, which enforces that $P(\bm{z}|\bm{y}, A=0)=P(\bm{z}|\bm{y}, A=1)$. By incorporating note labels as an extra input to the attribute predictor $\psi$, we can eliminate the conditional dependencies between the acoustic features and the gender labels. Since the note labels are available for both demographic groups, we propose two variants for the design of condition. In variant 1, the ground-truth notes $\bm{y}$ are transformed into frame-level labels $\bm{O}, \bm{S}, \bm{V}, \bm{P}$, which are then fed into the attribute predictor. For variant 2, the logits of frame-level predictions $\hat{O}_t, \hat{S}_t, \hat{V}_t, \hat{P}_t$ are used instead. 

From the perspective of fairness criteria, the variant 1 enforces $\bm{z}\perp A|\bm{y}$ while the variant 2 enforces $\bm{z}\perp A|\hat{\bm{y}}$. Therefore, the former can enforce $\hat{\bm{y}}\perp A|\bm{y}$, which is the separation criteria \cite{barocas-hardt-narayanan}. The formulation of variant 2 is motivated by the perspective of distribution alignment. The prediction $\hat{\bm{y}}$ contains prior knowledge about the classifier $\phi$ and similarities among different pitches (the predicted probabilities of other pitch values besides the true pitch value are not zeros, similar to dark knowledge in knowledge distillation \cite{hinton2015distilling}), while the ground truth $\bm{y}$ is one hot and has no information about other pitches. Considering that the pitch distributions of two gender groups are different, the proportion of male and female labels for each specific pitch value is also different. When conducting the conditional alignment for a specific pitch, the similarities contained in the pitch logits can assist in aligning other pitches, resulting in an improved and more efficient conditional alignment by the attribute predictor. 

The framework of note-conditioned adversarial learning for SVT is depicted in Fig. \ref{fig_framework}, and its variant 2 is presented in Alg. \ref{alg1}. The algorithm of variant 1 can be similarly derived. To train the SVT system, we follow the linear-probing and full-finetuning strategy proposed in \cite{gu2023deep}. During the linear probing stage, only the label predictor $\phi$ and attribute predictor $\psi$ are updated. This stage serves as a warm-up for these two models and preserves the pre-trained features of SSL models. Then in the full-finetuning stage, all three models $\theta, \phi, \psi$ are updated using different learning rates. To parameterize the note-conditioned attribute predictor, we first use two linear layers to embed each frame of acoustic features $\bm{z}$ and notes $\bm{y}$ (or $\hat{\bm{y}}$), respectively. Then the two embeddings are concatenated and passed through two linear layers with ReLU activations. We note that by modifying the loss function in Eq. \ref{eq6}, our bias mitigation framework can be readily applied to other types of sensitive attributes, such as $A$ is a multi-class attribute, or a continuous attribute, or a vector that encompasses multiple attributes. 

\subsection{Fairness-Utility Trade-off}\label{tradeoff}

Our task involves an inherent trade-off between fairness $F$ and utility $U$, as noted in previous literature \cite{kleinberg2016inherent}. This is due to the fact that when the gender information is removed from the acoustic features, the representations used for the SVT task may be affected. Optimizing $\mathcal{L}_y-\lambda\mathcal{L}_A$ of the acoustic encoder $\theta$ poses a challenge as the the two loss terms have conflicting natures. Empirical experiments demonstrate that in some cases, improvements in the fairness metric $F$ come at a cost of reduced utility metric $U$. Our goal is to improve the fairness metric $U$ without significantly degrading the utility metric $U$, starting from the initial point $(F_0, U_0)$ without mitigating bias. To achieve this, we introduce a tolerance hyper-parameter $\delta$ for $U$. We aim to increase the value of $F$ as much as possible within the range of $U>U_0-\delta$. Meanwhile, sacrificing utility beyond this range is not acceptable for real-world applications. This trade-off criterion facilitates the model selection. Typically, we set $\delta$ as $2\%$ or $5\%$ for f1-scores of COnPOff and COnP.

\begin{algorithm}[t]
\caption{Note-conditioned adversarial learning for SVT}\label{alg1}
\begin{algorithmic}
\Require Acoustic encoder $\theta^{(0)}$ pre-trained on speech data under SSL objective, randomly initialized label predictor $\phi^{(0)}$ and attribute predictor $\psi^{(0)}$, learning rates $\eta_1, \eta_2, \eta_3$ for $\theta, \phi, \psi$, training steps $K_1, K_2$ for linear probing and full finetuning.
\For{$k=1$ \textbf{to} $K_1+K_2$}
\State $\bm{z}=\theta^{(k-1)}(\bm{x})$, $\hat{\bm{y}}=\phi^{(k-1)}(\bm{z})$
\State $\hat{A}_t=\psi^{(k-1)}(\bm{z}\text{.detach}(), \hat{\bm{y}}\text{.detach}())$, compute $\mathcal{L}_A$ in Eq. \ref{eq6}
\State $\psi^{(k)}=\psi^{(k-1)}-\eta_3\frac{\partial\mathcal{L}_{A}}{\partial\psi^{(k-1)}}$ \Comment{Update $\psi$}
\State $\hat{A}_t=\psi^{(k)}(\bm{z}, \hat{\bm{y}})$, compute $\mathcal{L}_y, \mathcal{L}_A$ in Eq. \ref{eq2} and \ref{eq6}
\State $\phi^{(k)}=\phi^{(k-1)}-\eta_2\frac{\partial\mathcal{L}_{y}}{\partial\phi^{(k-1)}}$ \Comment{Update $\phi$}
\If{$k\leq K_1$}  \Comment{Update $\theta$}
\State $\theta^{(k)}=\theta^{(k-1)}$
\Else
\State $\theta^{(k)}=\theta^{(k-1)}-\eta_1(\frac{\partial\mathcal{L}_{y}}{\partial\theta^{(k-1)}}-\lambda\frac{\partial\mathcal{L}_{A}}{\partial\theta^{(k-1)}})$
\EndIf
\EndFor
\end{algorithmic}
\end{algorithm}

\begin{table*}[t]
\caption{Bias mitigation performance of note-conditioned adversarial learning.}
\begin{tabular}{l|l|l|ll|ll}
\toprule
\multirow{2}{*}{Train set} & \multirow{2}{*}{Test set} &\multirow{2}{*}{Method} & \multicolumn{2}{c}{COnPOff (\%)} & \multicolumn{2}{|c}{COnP (\%)} \\ \cline{4-7} 
& & & Utility ($U$) $\uparrow$    & Fairness ($F$) $\uparrow$   & Utility ($U$) $\uparrow$    & Fairness ($F$) $\uparrow$    \\
\midrule
\multirow{2}{*}{M4Singer} & \multirow{2}{*}{M4singer} & ERM      & 53.66   & -\,\,7.34    & 61.95   & -\,\,8.98   \\
&&Ours     & 52.48 (\textcolor{red}{-1.18})   & -\,\,3.61 (\textcolor{green}{+3.73})    & 60.67 (\textcolor{red}{-1.28})   & -\,\,4.21 (\textcolor{green}{+4.77})  \\
\midrule
\multirow{4}{*}{MIR-ST500} & \multirow{2}{*}{N20EMv2} & ERM      & 55.20   & -\,\,9.41    & 72.03   & -12.65   \\
&&Ours     & 53.29 (\textcolor{red}{-1.91})   & -\,\,4.14 (\textcolor{green}{+5.27})    & 72.71 (\textcolor{green}{+0.68})   & -10.82 (\textcolor{green}{+1.83})  \\
\cline{2-7} 
& \multirow{2}{*}{ISMIR2014} & ERM      & 52.58   & -15.94    & 67.75   & -13.20   \\
&&Ours     & 48.18 (\textcolor{red}{-4.40})   & -\,\,8.29 (\textcolor{green}{+7.65})    & 65.40 (\textcolor{red}{-2.35})   & -\,\,9.08 (\textcolor{green}{+4.12})  \\
\midrule
 & \multirow{2}{*}{N20EMv2} & ERM      & 73.06   & -\,\,8.65    & 79.56   & -\,\,7.89   \\
MIR-ST500 &&Ours     & 72.43 (\textcolor{red}{-0.63})   & -\,\,5.78 (\textcolor{green}{+2.87})    & 78.47 (\textcolor{red}{-1.09})   & -\,\,6.82 (\textcolor{green}{+1.07})  \\
\cline{2-7} 
N20EMv2 & \multirow{2}{*}{ISMIR2014} & ERM      & 59.95   & -10.45    & 73.85   & -12.36   \\
&&Ours     & 59.57 (\textcolor{red}{-0.38})   & -\,\,7.49 (\textcolor{green}{+2.96})    & 73.33 (\textcolor{red}{-0.52})   & -\,\,7.75 (\textcolor{green}{+4.61})  \\
\bottomrule
\end{tabular}
\label{tbl_cdann}
\vskip -0.1 in
\end{table*}

\section{Empirical Experiments}
\subsection{Bias Mitigation Performance}
To evaluate in-domain fairness, we adopt the proposed note-condi- tioned adversarial learning method on the M4Singer dataset. The SVT system is based on wav2vec 2.0 \cite{baevski2020wav2vec} and trained on the training split of M4Singer and evaluated on its test split. We set the learning rates for acoustic encoder and label predictor to be fixed at $\eta_1=\eta_2=3\times10^{-4}$. To further evaluate out-of-domain fairness, we conduct experiments on model1 and model3, as displayed in Table \ref{tbl_bias}. We set the learning rates to fixed values of $\eta_1=5\times10^{-5}, \eta_2=3\times10^{-4}$ following \cite{gu2023deep}. During the bias mitigation, we select the learning rate for the attribute predictor $\eta_3$ from the set $\{0.1, 0.01, 0.001, 0.0001\}$ and the balancing term $\lambda$ from the set $\{0.2, 0.5, 1.0, 2.0\}$. We report the best results we can achieve. This hyper-parameter selection also applies to the baselines we compare with in the following Sec. \ref{ablation}. We find that further increasing the learning rate $\eta_3$ or the balancing term $\lambda$ results in severe degradation in utility, even though the gender bias seems to be eliminated, as elaborated in Sec. \ref{failure}. Additionally, when $\eta_1$ is large, we find that the best results are achieved when $\eta_3$ is also large. Given our framework is based on adversarial learning, aiming to achieve equilibrium between the acoustic encoder and the attribute predictor, a larger learning rate for the encoder necessitates a corresponding increase in the learning rate of the attribute predictor to attain equilibrium\footnote{We conducted our experiments using the open-sourced repo: \url{https://github.com/guxm2021/SVT_SpeechBrain}.}.

Table \ref{tbl_cdann} presents a comparison between the SVT systems trained with our note-conditioned adversarial learning (variant 2) and those trained using empirical risk minimization (ERM) in Eq. \ref{eq2} without bias mitigation. Our experiments on the M4Singer dataset reveal that the fairness metrics improve by over 50\% for COnPOff and COnP, respectively. At the same time, the utility metrics drop only by 1.18\% and 1.28\% for COnPOff and COnP, respectively. Apart from the in-domain fairness results, we evaluate the out-of-domain fairness results on model1 and model3 (mentioned in Sec. \ref{gap_data}). Our experiments demonstrate that applying the bias mitigation method on model1 significantly improves its fairness. In particular, the performance disparity decreases by $50\%$ on N20EMv2 and ISMIR2014, in terms of COnPOff. These results validate the effectiveness of our bias mitigation method on out-of-domain data. Furthermore, we achieve fairer SVT performance on both in-domain and out-of-domain scenarios with minor total performance degradation when using the state-of-the-art performing SVT system (model3). 

\begin{table*}[h]
\caption{More bias mitigation results.}
\begin{tabular}{l|l|l|c|c|ll|ll}
\toprule
\multirow{2}{*}{Train set} & \multirow{2}{*}{Test set} &\multirow{2}{*}{Method} & \multirow{2}{*}{$\eta_3$} & \multirow{2}{*}{$\lambda$} &  \multicolumn{2}{c}{COnPOff (\%)} & \multicolumn{2}{|c}{COnP (\%)} \\ \cline{6-9} 
& & & & & Utility ($U$) $\uparrow$    & Fairness ($F$) $\uparrow$   & Utility ($U$) $\uparrow$    & Fairness ($F$) $\uparrow$    \\
\midrule
\multirow{4}{*}{M4Singer} & \multirow{4}{*}{M4singer} & ERM      & - & - &  53.66   & -\,\,\,7.34   & 61.95   & -\,\,\,8.98  \\
&&AL     & 1.0 & 3.0 & 42.56 (\textcolor{red}{-11.10})   & +\,\,4.65    & 49.60 (\textcolor{red}{-12.35})   & +\,\,6.35 \\
&&NCAL     & 0.1 & 5.0 & 49.97 (\textcolor{red}{-\,\,3.69})   & +\,\,0.39    & 57.66 (\textcolor{red}{-\,\,4.29})   & +\,\,1.57 \\
&&NCAL     & 0.1 & 3.0 & 49.87 (\textcolor{red}{-\,\,3.79})   & +\,\,1.29    & 58.53 (\textcolor{red}{-\,\,3.42})   & +\,\,0.60 \\
\midrule
\multirow{4}{*}{MIR-ST500} & \multirow{2}{*}{N20EMv2} & ERM  & - & -   & 55.20   & -\,\,\,9.41    & 72.03   & -12.65   \\
&&NCAL   & 0.0001 & 5.0    & 45.78 (\textcolor{red}{-\,\,9.42})   & +\,\,2.45   & 69.54 (\textcolor{red}{-\,\,2.49})   & -\,\,\,3.77  \\
\cline{2-9} 
& \multirow{2}{*}{ISMIR2014} & ERM   & - & -    & 52.58   & -15.94    & 67.75   & -13.20   \\
&&NCAL  & 0.0001 & 5.0   & 42.24 (\textcolor{red}{-10.34})   & +\,\,3.01    & 62.44 (\textcolor{red}{-\,\,5.31})   & +\,\,1.22  \\
\midrule
 & \multirow{2}{*}{N20EMv2} & ERM  & - & -     & 73.06   & -\,\,\,8.65    & 79.56   & -\,\,\,7.89   \\
MIR-ST500 &&NACL & 0.001 & 5.0    & 72.28 (\textcolor{red}{-\,\,0.78})   & -\,\,\,7.78    & 78.70 (\textcolor{red}{-\,\,0.86})   & -\,\,\,8.23  \\
\cline{2-9} 
N20EMv2 & \multirow{2}{*}{ISMIR2014} & ERM  & - & -     & 59.95   & -10.45    & 73.85   & -12.36   \\
&&NACL  & 0.001 & 5.0   & 50.95 (\textcolor{red}{-\,\,9.00})   &+\,\,2.49  & 65.87 (\textcolor{red}{-\,\,7.98})   & +\,\,0.64  \\
\bottomrule
\end{tabular}
\label{tbl_failure}
\vskip -0.1 in
\end{table*}

\subsection{Comparisons with Baselines}\label{ablation}
We compare our note-conditioned adversarial learning framework with two baseline methods: adversarial learning (AL), and domain independent training proposed in \cite{wang2020towards}. We denote our method as ``NCAL (variant 1)'' and ``NCAL (variant 2)''. For AL, we keep the same configuration as our NCAL, except that we do not feed any condition into the attribute predictor. For domain independent training, we compare our method with two variants: calibrated inference and miscalibrated inference, which differ in whether the gender labels are used during the inference. These two inferences are abbreviated as `` DInD (w/ calibr.)'' and ``DInD (w/o calibr.)'', respectively. We refer readers to \cite{wang2020towards} for more technical details. We evaluate the comparisons on model3 and report the results on the N20EMv2 and ISMIR2014 datasets. 

\begin{figure}[h]
\centering
\includegraphics[width=\linewidth]{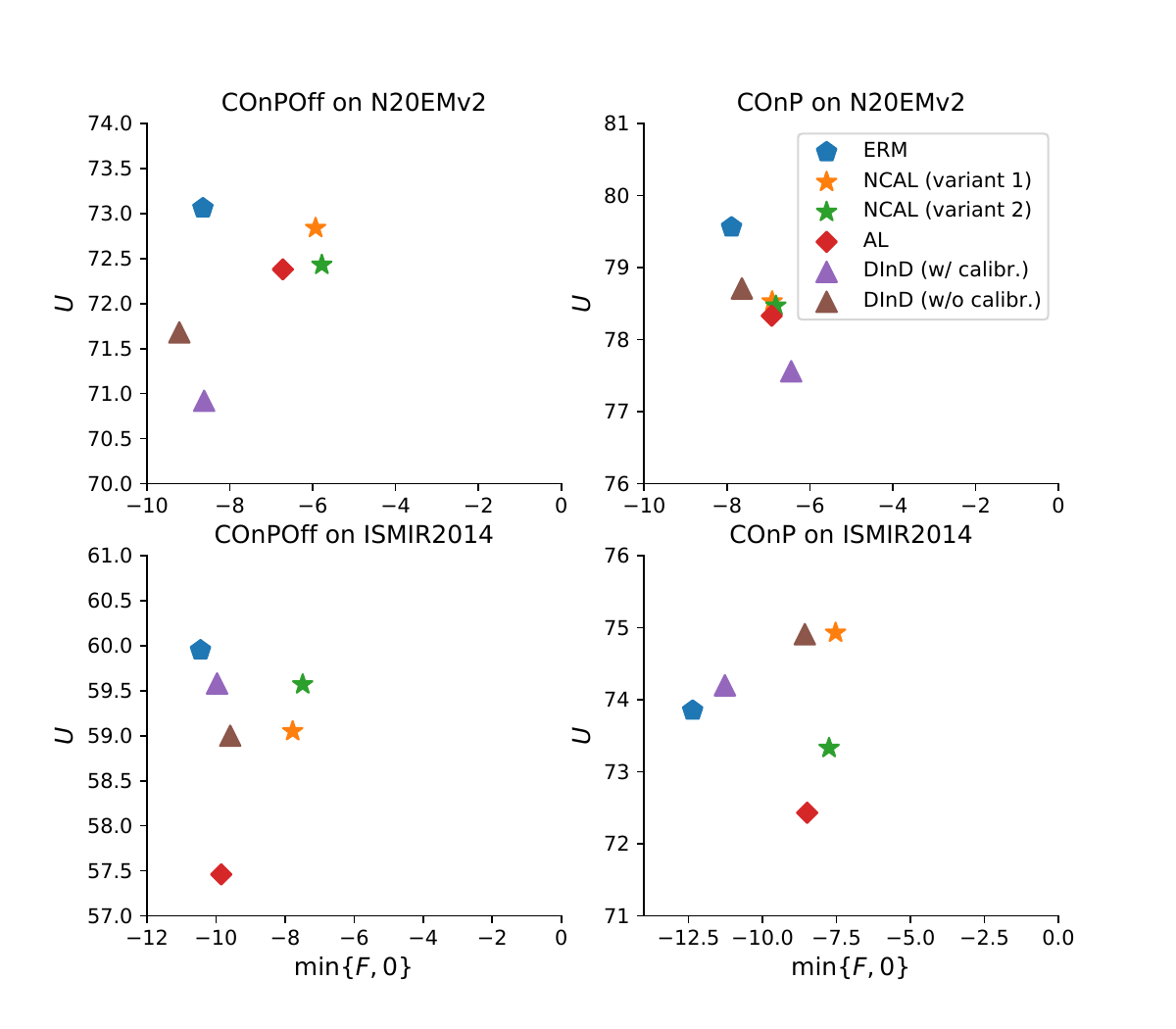}
\caption{Comparisons among different bias mitigation methods of Fairness-Utility trade-off on N20EMv2 and ISMIR2014.}
\label{fig_ablation}
\vskip -0.2 in
\end{figure}

We present the Fairness-Utility trade-off of various bias mitigation methods in Fig. \ref{fig_ablation}. According to the optimization bundle in Eq. \ref{eq5}, an upper-right point signifies a better trade-off compared to a lower-left point. Firstly, we observe that our two variants of NACL, ``NCAL (variant 1)'' and ``NCAL (variant 2)'', perform the best in the most cases in terms of fairness-utility trade-off. These two variants perform similarly on N20EMv2. On the ISMIR2014 dataset, NACL variant 2 performs better for COnPOff while variant 1 exhibits superiority for COnP. We then note that NCAL consistently outperforms the baseline AL. The performance of AL is similar to NACL only in terms of COnP on N20EMv2. However, in other cases, AL shows lower utility and less fairness. With respect to DInD, it shows better fairness than NCAL only in terms of COnP on N20EMv2 with calibration. However, its utility is more severely degraded than NCAL. In other cases, NACL consistently outperforms DInD. Although adversarial learning approaches are prone to instability during training compared to non-adversrial learning approaches, such as DInD, our proposed NCAL offers a better fairness-utility trade-off. Moreover, DInD cannot be easily applied to the scenarios where continuous sensitive attributes or multiple sensitive attributes are considered. 

\subsection{Further Empirical Analysis}\label{failure}
In our experiments, we observe that setting a larger learning rate $\eta_3$ or the balancing term $\lambda$ leads to near-perfect gender fairness but a drastic degradation in utility. As presented in Table \ref{tbl_failure}, on M4Singer, our NCAL method can achieve almost gender performance equality when $\lambda$ is set to 3.0 or 5.0, but at the cost of more utility degradation than the results in Table \ref{tbl_cdann}. However, the baseline AL method fails to achieve such equality. We only observe a better performance for the male group with much more utility deterioration compared to NCAL when $\eta_3=1.0$ and $\lambda=3.0$. This trend is consistently observed on model1 and model3 using NCAL for bias mitigation. We hypothesize that increasing either $\eta_3$ or $\lambda$ enhances the discrimination ability of the attribute predictor, causing the models to focus more on the fairness while neglecting the utility. As explained in Sec. \ref{tradeoff}, such models may not be suitable for real-world applications where both fairness and utility are important.

The results in Table \ref{tbl_cdann} and \ref{tbl_failure} can also validate our assumptions behind the performance disparity in SVT. Our baseline AL aims to ensure that the learnt acoustic representations contains as little gender information as possible. In this way, the effects of sound characteristics, which are related to gender, can be implicitly mitigated. The improvements in terms of fairness metrics brought by baseline AL provide evidence about our assumption that gender bias in SVT performance is attributed to the differences of sound characterises across different demographic groups. Additionally, when conditioning on the pitch information, our SVT systems could achieve nearly perfect fairness, as presented in Table \ref{tbl_failure}, which further demonstrates the substantial contributions of pitch distribution difference to the performance disparity.

\section{Discussion and Future Work}
In addition to the primary focus on group fairness in this work, max-min fairness \cite{pham2023fairness} is also raised in certain cases. Max-min fairness aims to minimize the worse-case error rates, offering an alternative perspective on fairness evaluation. Our motivation stems from the objective of enhancing user experience and mitigating potential discriminatory treatment when utilizing SVT systems and their downstream applications. Consequently, our main target is to strive for performance equalization across different demographic groups. Therefore, there is less discussion on max-min fairness. Despite this, we still observed improved performance for male data in most cases after applying our bias mitigation approach. 

In this work, we formulate our solution from the perspective of fairness. We think the perspective of signal processing may also be beneficial to further interpret our findings. By incorporating signal processing approaches, our adversarial learning framework may further enhance SVT performance in terms of both fairness and utility. We identify this as an avenue for future exploration. Furthermore, we think our approach could be extended to consider other sensitive attributes, such as age, race, language. Beyond demographic groups, we also recognize that different instruments generally exhibit distinct pitch distributions and timbre characteristics. Hence, our approach holds potential applicability in the domain of automatic music transcription \cite{gardner2021mt3}, wherein musical notes are inferred from audio signals produced by diverse instruments.

\section{Conclusion}
This work represents the first attempt of fairness topic within the singing-centric deep learning community. We presented evidence that the performance of singing voice transcription (SVT) on female data surpasses that of male data, irrespective of the models or datasets employed. Our findings suggested that this performance disparity is attributed to the inherent differences between male and female singing voices, especially in pitch distribution. Given the significance of this fairness issue, we proposed a note-conditioned adversarial learning approach to mitigate gender bias in SVT. Specifically, our approach leveraged an attribute predictor to learn gender-invariant acoustic representations. By conditioning on the note events, we further achieved conditional alignment between acoustic features of different groups. Our results demonstrated the effectiveness of our bias mitigation method, as it significantly improves fairness metrics while maintaining utility metrics across both in-domain and out-of-domain data. 
\begin{acks}
We would like to thank anonymous reviewers for their valuable suggestions. We also appreciated Nicholas Wong's writing advice and Xudong Shen's comments on fairness. This project is funded in part by a research grant MOESOL-2021-0017 from the Ministry of Education in Singapore.
\end{acks}
\clearpage
\bibliographystyle{ACM-Reference-Format}
\balance
\bibliography{fair}

\end{document}